\documentclass[11pt,a4paper]{article}
\usepackage{amssymb}
\usepackage{epsf}
\renewcommand{\a}{\textsl{a}}
\begin{document}

\title{Quantum Field Theory in Curved Spacetime\footnote
{Published as {\sl Encyclopedia of Mathematical Physics}, J.-P. 
Fran\c coise, G. Naber and T.S. Tsou, eds., Academic (Elsevier),
Amsterdam, New York, London 2006, Vol. 4, pp 202-212}}
\author{Bernard S.~Kay \\
\normalsize Department of Mathematics,
University of York, York YO10 5DD, U.K.}

\maketitle

\section{Introduction and Preliminaries}

Quantum Field Theory (QFT) in Curved Spacetime is a hybrid approximate
theory in which quantum matter fields are assumed to propagate in a
fixed classical background gravitational field.  Its basic physical
prediction is that strong gravitational fields can polarize the vacuum
and, when time-dependent, lead to pair creation just as a strong and/or
time-dependent electromagnetic field can polarize the vacuum and/or give
rise to pair-creation of charged particles.  One expects it to be a good
approximation to full quantum gravity provided the typical frequencies
of the gravitational background are very much less than the Planck
frequency ($c^5/G\hbar)^{1/2}\sim 10^{43}s^{-1}$) and provided, with a
suitable measure for energy, the energy of created particles is very
much less than the energy of the background gravitational field or of
its matter sources. Undoubtedly the most important prediction of the
theory is the Hawking effect, according to which a, say spherically
symmetric, classical black hole of mass $M$ will emit thermal radiation
at the Hawking temperature  $T=(8\pi M)^{-1}$  (here and from now on, we
use Planck units  where $G$, $c$, $\hbar$ and $k$ [Boltzmann's constant]
are all taken to be 1.)

On the mathematical side, the need to formulate the laws and derive the
general properties of  QFT on non-flat spacetimes forces one to state
and prove results in local terms and, as a byproduct, thereby leads to
an improved perspective on flat-space-time QFT too.  It's also
interesting  to formulate QFT on idealized spacetimes with particular
global geometrical features.   Thus, QFT on spacetimes with bifurcate
Killing horizons is intimately related  to the Hawking effect; QFT on
spacetimes with closed timelike curves is intimately related to the
question whether the laws of physics permit the manufacture of a
time-machine.

As is standard in General Relativity, a curved spacetime is modelled
mathematically as a  (paracompact, Hausdorff) manifold $\cal M$ equipped
with a pseudo-Riemannian metric $g$ of signature $(-,+++)$ (we follow
the conventions of the standard text ``{\it Gravitation}'' by Misner
Thorne and Wheeler [W.H. Freeman, San Francisco, 1973]).   We shall also
assume, except where otherwise stated, our spacetime to be {\it globally
hyperbolic}, i.e. that $\cal M$ admits a {\it global time coordinate},
by which we mean a global coordinate $t$ such that each constant-$t$
surface is a smooth Cauchy surface  i.e. a smooth spacelike 3-surface
cut exactly once by each inextendible causal curve.  (Without this
default assumption, extra problems arise for QFT which we shall briefly
mention in connection with the time-machine question in  Section 6.) In
view of this definition, globally hyperbolic spacetimes are clearly
time-orientable and we shall assume a choice of time-orientation has
been made  so we can talk about the ``future'' and ``past'' directions.
Modern formulations of the subject take, as the fundamental mathematical
structure modelling the quantum field, a $*$-algebra $\cal A$ (with
identity $I$) together with a family of local sub $*$-algebras ${\cal
A}({\cal O})$ labelled by bounded open regions $\cal O$ of the spacetime
$({\cal M},g)$ and satisfying the {\it isotony} or {\it net} condition
that ${\cal O}_1 \subset {\cal O}_2$ implies ${\cal A}({\cal O}_1)$ is a
subalgebra of  ${\cal A}({\cal O}_2)$, as well as the condition that whenever
${\cal O}_1$ and ${\cal O}_2$ are spacelike separated, then ${\cal
A}({\cal O}_1)$ and ${\cal A}({\cal O}_2)$ commute.

Standard concepts and techniques from algebraic quantum theory are then
applicable:  In particular, {\it states} are defined to be positive
(this means $\omega(A^*A) \ge 0$ $\forall A  \in {\cal A}$) normalized
(this means $\omega(I)=1$) linear functionals on $\cal A$.  One
distinguishes between {\it pure} states and {\it mixed} states, only the
latter being writeable as non-trivial convex  combinations of other
states. To each state, $\omega$, the {\it GNS-construction} associates a
representation, $\rho_\omega$,  of $\cal A$ on a Hilbert space ${\cal
H}_\omega$ together with a cyclic vector  $\Omega\in{\cal H}_\omega$
such that
$$\omega (A)=\langle\Omega|\rho_\omega(A)\Omega\rangle$$
(and the {\it GNS triple} $(\rho_\omega,{\cal H},\Omega)$ is unique up
to equivalence). There are often technical advantages in formulating
things so that the $*$-algebra is a C$^*$-algebra.  Then the GNS
representation is as everywhere-defined bounded operators and  is
irreducible if and only if the state is pure.  A useful concept, due to
Haag, is the {\it folium} of a given state $\omega$ which may be defined
to be the set of all states $\omega_\sigma$ which arise in the form
$\rm{Tr}(\sigma\rho_\omega(\cdot))$ where $\sigma$ ranges over the
density operators (trace-class operators with unit trace) on ${\cal
H}_\omega$.

Given a state, $\omega$, and an automorphism, $\alpha$, which preserves
the state (i.e. $\omega\circ\alpha=\omega$) then there will be a unitary
operator, $U$, on $H_\omega$ which {\it implements} $\alpha$ in the
sense that  $\rho_\omega(\alpha(A))=U^{-1}\rho_\omega(A)U$ and $U$ is
chosen uniquely by the condition $U\Omega=\Omega$.

On a {\it stationary} spacetime, i.e. one which admits a one-parameter
group of isometries whose integral curves are everywhere timelike,  the
algebra will inherit a one-parameter group (i.e. satisfying
$\alpha(t_1)\circ\alpha(t_2)=\alpha(t_1+t_2)$) of time-translation
automorphisms, $\alpha(t)$, and, given any stationary state (i.e. one
which satisfies $\omega\circ\alpha(t)=\omega\quad \forall t\in {\bf R}$)
these will be implemented by a one-parameter group of unitaries, $U(t)$,
on its GNS Hilbert space satsifying $U(t)\Omega=\Omega$.  If $U(t)$ is
strongly continuous so that it takes the form $e^{-iHt}$ and if the
Hamiltonian, $H$, is positive, then $\omega$ is said to be a {\it ground
state}.  Typically one expects ground states to exist and often be
unique.

Another important class of stationary states for the algebra of a
stationary spacetime is the class of {\it KMS states}, $\omega^\beta$,
at inverse temperature $\beta$; these have the physical interpretation
of thermal equilibrium states. In the GNS representation of one of
these,  the automorphisms are also implemented by a strongly-continuous
unitary group, $e^{-iHt}$, which preserves $\Omega$ but (in place of $H$
positive) there is a complex conjugation, $J$, on $H_\omega$ such that
\begin{equation} \label{kms}
e^{-\beta H/2}\rho_\omega(A)\Omega=J\rho_\omega(A^*)\Omega
\end{equation}for all $A\in {\cal A}$.
An attractive feature of the subject is that its main qualitative
features are already present for linear field theories and, unusually in
comparison with other questions in QFT, these are susceptible of a
straightforward explicit and rigorous mathematical formulation.  In
fact, as our principal example, we give, in Section 2 a construction for
the field algebra for the quantized real linear Klein Gordon equation
\begin{equation} (\Box_g-m^2-V)\phi=0 \label{kg} \end{equation} of mass
$m$ on a globally hyperbolic spacetime $({\cal M},g)$. Here, $\Box_g$
denotes the Laplace Beltrami operator $g^{ab}\nabla_a\partial_b$
($=(|\det(g)|)^{-1/2}\partial_a((|\det(g)|^{1/2}g^{ab}\partial_b)$).  We
include a scalar external background classical field, $V$ in addition to
the external gravitational field represented by $g$.  In case $m$ is
zero, taking $V$ to equal $R/6$ where $R$ denotes the Riemann scalar,
makes the equation {\it conformally invariant}.

The main new feature of quantum field theory in curved spacetime
(present already for linear field theories) is that, in a general
(neither flat, nor stationary) spacetime there will not be any single
preferred state but rather a family of preferred states, members of
which are best regarded  as on an equal footing with one-another. It is
this feature which makes the above algebraic framework particularly
suitable,  indeed essential, to a clear formulation of the subject.
Conceptually it is this feature which takes the most getting used to.
In particular, one must realize that, as we shall explain in Section 3,
the interpretation of a state as having a particular
``particle-content'' is in general problematic because it can only be
relative to a particular choice of ``vacuum'' state and, depending on
the spacetime of interest, there may be one state or several states  or,
frequently, no states at all which deserve the name ``vacuum'' and even
when  there are states which deserve this name, they will often only be
defined in some approximate or asymptotic or transient sense or only on
some subregion of the spacetime.

Concomitantly, one does not expect global observables such as the
``particle number'' or the quantum Hamiltonian of flat-spacetime free
field theory to generalize to a curved spacetime context, and for this
reason local observables play a central role in the theory.  The
quantized stress-energy tensor is a particularly natural and important
such local observable and the theory of this is central to the whole
subject.  A brief introduction to it is given in Section 4.

This is followed by a further Section 5 on the Hawking and Unruh effects
and a brief Section 6 on the problems of extending the theory beyond the
``default'' setting,  to non-globally hyperbolic spacetimes.  Finally,
Section 7 briefly mentions a number of other interesting and active
areas of the subject as well as issuing a few {\it warnings} to be borne
in mind when reading the literature.

\section{Construction of $*$-Algebra(s) for a Real Linear Scalar Field on
Globally Hyperbolic Spacetimes and Some General Theorems}

On a globally hyperbolic spacetime, the classical equation (\ref{kg})
admits well-defined {\it advanced} and {\it retarded Green} functions
(strictly bidistributions) $\Delta^A$ and $\Delta^R$ and the standard
covariant quantum free real (or ``Hermitian'') scalar field commutation
relations familiar from Minkowski spacetime free field theory naturally
generalize to the (heuristic) equation
$$[\hat\phi(x), \hat\phi(y)]=i\Delta(x,y)I$$
where $\Delta$ is the {\it Lichn\'erowicz commutator function}
$\Delta=\Delta^A-\Delta^R$.  Here, the ``$\hat{\quad}$'' on the quantum
field $\hat\phi$ serves to distinguish it from a classical solution
$\phi$. In mathematical work, one does not assign a meaning to the field
at a point itself, but rather aims to assign meaning to {\it smeared
fields} $\hat\phi(F)$ for all real-valued test functions $F\in
C_0^\infty({\cal M})$ which are then to be interpreted as standing for
$\int_{\cal M}\hat\phi(x)f(x)|\det(g)|^{1\over 2}d^4x$. In fact, it is
straightforward to   define a {\it minimal field algebra} (see below)
${\cal A}_{\hbox{min}}$ generated by such $\hat\phi(F)$ which satisfy
the suitably smeared version
$$[\hat\phi(F), \hat\phi(G)]=i\Delta(F,G)I$$
of the above commutation relations together with  Hermiticity (i.e.
$\hat\phi(F)^*=\hat\phi(F)$), the property of being a weak solution of
the equation (\ref{kg}) (i.e.    $\hat\phi((\Box_g-m^2-V)F)=0\quad
\forall F\in C_0^\infty({\cal M})$) and linearity in test functions.
There is a technically different alternative formulation of this minimal
algebra, which is known as the {\it Weyl algebra}, which is constructed
to be the C$^*$ algebra generated by operators $W(F)$ (to be interpreted
as standing for $\exp(i\int_{\cal M}\hat\phi(x)f(x)|\det(g)|^{1\over
2}d^4x$) satisfying
$$W(F_1)W(F_2)=\exp(-i\Delta(F_1,F_2)/2)W(F_1+F_2)$$
together with $W(F)^*=W(-F)$ and $W((\Box_g-m^2-V)F)=I$.  With either
the minimal algebra or the Weyl algebra one can define, for each bounded
open region $\cal O$, subalgebras $A({\cal O})$ as generated by the
$\hat\phi(\cdot)$ (or the $W(\cdot)$) smeared with test functions
suppported in ${\cal O}$ and verify that they satisfy the above ``net''
condition and commutativity at spacelike separation.

Specifying a state, $\omega$, on ${\cal A}_{\hbox{min}}$  is tantamount
to specifying its collection of {$n$-point distributions} (i.e. smeared
$n$-point functions) $\omega(\hat\phi(F_1)\dots\hat\phi(F_n))$ (In the
case of the Weyl algebra, one restricts attention to ``regular'' states
for which the map $F\rightarrow \omega(W(F))$ is sufficiently often
differentiable on finite dimensional subspaces of $C_0^\infty({\cal M})$
and defines the n-point  distributions in terms of derivatives with
respect to suitable parameters of expectation values of suitable Weyl
algebra elements.)  A particular role is played in the theory by the
{\it quasi-free} states for which all the {\it truncated} $n$-point
distributions except for $n=2$ vanish.  Thus all the n-point
distributions for odd $n$ vanish while the 4-point distribution is made
out of the 2-point distribution according to
$$\omega(\hat\phi(F_1)\hat\phi(F_2)\hat\phi(F_3)\hat\phi(F_4))=
\omega(\hat\phi(F_1)\hat\phi(F_2))\omega(\hat\phi(F_3)\hat\phi(F_4))$$
$$+\omega(\hat\phi(F_1)\hat\phi(F_3))\omega(\hat\phi(F_2)\hat\phi(F_4))
+ \omega(\hat\phi(F_1)\hat\phi(F_4))\omega(\hat\phi(F_2)\hat\phi(F_3))$$
etc. The anticommutator distribution
\begin{equation} \label{anticom}
G(F_1,F_2) =
\omega(\hat\phi(F_1)\hat\phi(F_2))+\omega(\hat\phi(F_2)\hat\phi(F_1))
\end{equation}
of a quasi-free state (or indeed of any state) will satisfy the properties
(for all test functions $F$, $F_1$, $F_2$ etc.):

\begin{description}

\item{(a)} {\it (symmetry)} $G(F_1,F_2)=G(F_2,F_1)$

\item{(b)} {\it (weak bisolution property)}
$$G((\Box_g - m^2 - V)F_1,F_2)=0=G(F_1,(\Box_g - m^2 - V)F_2)$$

\item{(c)} {\it (positivity)} $G(F,F)\ge0$ and
$G(F_1,F_1)^{1/2}G(F_2,F_2)^{1/2}\ge |\Delta(F_1,F_2)|$

\end{description}

\noindent
and it can be shown that, to every bilinear functional $G$ on
$C_0^\infty({\cal M})$ satisfying (a), (b) and (c), there is a
quasi-free state with two-point distribution $(1/2)(G + i\Delta)$. One
further declares a quasi-free state to be {\it physically admissible}
only if (for pairs of points in sufficiently small convex
neighbourhoods)

\begin{description}

\item{(d)} {\it (Hadamard condition)} ``$G(x_1,x_2)={1\over 2\pi^2}\left
(u(x_1,x_2){\rm P}{1\over\sigma}+v(x_1,x_2)\log|\sigma|+w(x_1,x_2)\right )$''

\end{description}

\noindent
This last condition expresses the requirement that (locally) the
two-point distribution actually ``is'' (in the usual sense in which one
says that a distribution ``is'' a function) a smooth function for pairs
of non-null-separated points, and at the same time requires that the
two-point distribution be singular at pairs of null-separated points and
locally specifies the nature of the singularity for such pairs of points
 with a leading ``principal part of $1/\sigma$'' type singularity and a
subleading ``$\log|\sigma|$'' singularity where $\sigma$ denotes the
square of the geodesic distance between $x_1$ and $x_2$. $u$ (which
satisfies $u(x_1,x_2)=1$ when $x_1=x_2$) and $v$ are certain smooth
two-point functions determined in terms of the local geometry and the
local values of $V$ by something called the {\it Hadamard procedure}
while the smooth two-point function $w$ depends on the state. We shall
omit the details.  The  important point is that this Hadamard condition
on the two-point distribution is believed to be the correct
generalization to a  curved-spacetime of the well-known universal
short-distance behaviour  shared by the truncated two-point
distributions of all physically relevant  states for the special case of
our theory when the spacetime is flat (and $V$ vanishes).  In the latter
case, $u$ reduces to $1$, and $v$ to a simple power series
$\sum_{n=0}^\infty v_n\sigma^n$ with $v_0=m^2/4$ etc.

Actually, it is known (this is the content of ``Kay's Conjecture'' which
was proved by M. Radzikowski in 1992) that (a), (b), (c) and (d)
together imply that the two-point distribution is nonsingular at all
pairs of (not necessarily close-together) spacelike-separated points.
More important than this result itself is a reformulation of the
Hadamard condition in terms of the concepts of {\it microlocal analysis}
which Radzikowski originally introduced as a tool towards its proof.

\begin{description}

\item{(d')} {\it (Wave Front Set [or Microlocal] Spectrum Condition)}
WF$(G+i\Delta)= \lbrace (x_1,p_1;x_2,p_2)\in
T^*({\cal M}\times {\cal M})\diagdown{\bf 0}|
x_1$ and $x_2$ lie on a single null geodesic, $p_1$ is tangent to
that null geodesic and future pointing, and $p_2$ when parallel
transported along that null geodesic from $x_2$ to $x_1$ equals
$-p_1\rbrace$

\end{description}

\noindent
For the gist of what this means, it suffices to know that to say that an
element $(x,p)$ of the cotangent bundle of a manifold (excluding the
zero section ${\bf 0}$) is in the {\it wave front set}, WF, of a given
distribution on that manifold may be expressed informally by saying that
that distribution is singular {\it at} the point $x$ {\it in the
direction} $p$.  (And here the notion is applied to $G+i\Delta$, thought
of as a distribution on the manifold ${\cal M}\times {\cal M}$.)

We remark that generically (and, e.g.,  always if the spatial sections are
compact and $m^2 + V(x)$ is everywhere positive) the Weyl algebra for equation
(\ref{kg}) on a given stationary spacetime will have a unique ground
state and unique KMS states at each temperature and these will be
quasi-free and Hadamard.

Quasi-free states are important also because of a theorem of R. Verch
(1994,  in verification of another conjecture of Kay) that (in the Weyl
algebra framework) on the algebra of any bounded open region, the folia
of the quasi-free Hadamard states coincide. With this result one can
extend the notion of physical admissibility to
not-necessarily-quasi-free states by demanding that, to be admissible, a
state belong to the resulting common folium when restricted to the
algebra of each bounded open region;  equivalently that it be a {\it
locally normal} state on the resulting natural extension of the net of
local Weyl algebras to a net of local {\it W$^*$ algebras}.

\section{Particle Creation and the Limitations of the Particle Concept}

Global hyperbolicity also entails that the Cauchy problem is well posed
for the classical field equation (\ref{kg}) in the sense that for every
Cauchy surface, ${\cal C}$, and every pair $(f,p)$ of Cauchy data in
$C_0^\infty({\cal C})$,  there exists a unique solution $\phi$ in
$C_0^\infty({\cal M})$ such that $f=\phi|_{\cal C}$ and
$p=|\det(g)|^{1/2}g^{0b}\partial_b\phi|_{\cal C}$.  Moreover $\phi$ has
compact support on all other Cauchy surfaces.  Given a global time
coordinate $t$, increasing towards the future, foliating $\cal M$ into a
family of contant-$t$ Cauchy surfaces, ${\cal C}_t$, and given a choice
of global time-like vector field $\tau^a$ (for example,
$\tau^a=g^{ab}\partial_b t$) enabling one to identify all the ${\cal
C}_t$, say with ${\cal C}_0$, by identifying points cut by the same
integral curve of $\tau^a$, a single such classical solution $\phi$ may
be pictured as a family $\lbrace (f_t, p_t): t \in {\bf R}\rbrace$  of
time-evolving Cauchy data on ${\cal C}_0$.  Moreover, since (\ref{kg})
implies, for each pair of classical solutions, $\phi_1$, $\phi_2$, the
conservation (i.e. $\partial_a j^a=0$) of the current
$j^a=|\det(g)|^{1/2}g^{ab}(\phi_1\partial_b\phi_2 -
\phi_2\partial_b\phi_1)$, the symplectic form 
(on $C_0^\infty({\cal C}_0)\times
C_0^\infty({\cal C}_0)$)
$$\sigma((f^1_t,p^1_t);(f^2_t,p^2_t))=
\int_{{\cal C}_0}(f^1_tp^2_t-p^1_tf^2_t)d^3x$$
will be conserved in time.  ($d^3x$ denotes $dx^1\wedge dx^2\wedge dx^3$.)

Corresponding to this picture of classical dynamics, one expects there
to be a description of quantum  dynamics in terms of a family of
sharp-time quantum fields $(\varphi_t, \pi_t)$ on ${\cal C}_0$,
satisfying heuristic canonical commutation relations
$$[\varphi_t({\bf x}),\varphi_t({\bf y})]=0, \quad [\pi_t({\bf x}),
\pi_t({\bf y})]=0, \quad
[\varphi_t({\bf x}), \pi_t({\bf y})]=i\delta^3({\bf x},{\bf y})I$$
and evolving in time according to the same dynamics as the Cauchy data
of a classical solution.  (Both these expectations are correct because
the field  equation is linear.)  An elegant way to make rigorous
mathematical sense of these expectations is in terms of a  $*$-algebra
with identity generated by Hermitian objects ``$\sigma((\varphi_0,
\pi_0); (f,p))$'' (``symplectically smeared sharp-time fields at
$t=0$'') satisfying linearity in $f$ and $p$ together with the
commutation relations
$$[\sigma((\varphi_0,\pi_0);(f^1,p^1)),\sigma((\varphi_0,\pi_0);(f^2,p^2))]
 =i\sigma((f^1,p^1);(f^2,p^2))I$$
and to define (symplectically smeared) time-$t$ sharp-time fields by demanding
$$\sigma((\varphi_t,\pi_t);(f_t,p_t))=\sigma((\varphi_0,\pi_0);(f_0,p_0))$$
where $(f_t,p_t)$ is the classical time-evolute of $(f_0,p_0)$. This
$*$-algebra of sharp-time fields may be identified with the (minimal)
field $*$-algebra of the previous section, the $\hat\phi(F)$ of the
previous section being identified with $\sigma((\varphi_0,\pi_0);(f,p))$
where $(f, p)$ are the Cauchy data at $t=0$ of $\Delta*F$.  (This
identification is of course many-one since $\hat\phi(F)=0$ whenever $F$
arises as $(\Box_g - m^2 - V)G$ for some test function $G\in
C_0^\infty({\cal M})$.)

Specializing momentarily to the case of the free scalar field $(\Box -
m^2)\phi=0$ ($m \ne 0$) in Minkowski space with a flat $t=0$ Cauchy
surface, the ``symplectically smeared'' two-point function of the usual
ground state (``Minkowski vacuum state''), $\omega_0$, is given, in this
formalism, by
\begin{equation} \label{mink}
\omega_0(\sigma((\varphi, \pi); (f^1, p^1))\sigma((\varphi, \pi); (f^2,
p^2)))={1\over 2}(\langle f^1|\mu f^2\rangle + \langle
p^1|\mu^{-1}p^2\rangle +i\sigma((f^1,p^1); (f^2,p^2))
\end{equation}
where the inner products are in the {\it one-particle Hilbert space}
${\cal H} = L^2_{\bf C}({\bf R}^3)$ and $\mu=(m^2-\nabla^2)^{1/2}$. The
GNS representation of this state may be concretely realised on the
familiar {\it Fock space} ${\cal F}({\cal H})$ over ${\cal H}$ by
$$\rho_0(\sigma((\varphi, \pi); (f,p)))= -i(\hat a^\dagger(\a)-(\hat
a^\dagger(\a))^*)$$ where $\a$ denotes the element of $\cal H$
$$\a={(\mu^{1/2}f+i\mu^{-1/2}p)\over\sqrt 2}$$
(we note in passing that, if we equip $\cal H$ with the symplectic
form $2\rm{Im}\langle\cdot|\cdot\rangle$, then $K: (f,p)\mapsto \a$ is a
symplectic map)
and $\hat a^\dagger(\a)$ is the usual smeared {\it creation operator}
(=``$\int \hat a^\dagger({\bf x})\a({\bf x})d^3{\bf x}$'') on
${\cal F}({\cal H})$ satisfying
$$[(\hat a^\dagger(\a^1))^*, \hat a^\dagger(\a^2)]=\langle
\a^1|\a^2\rangle_{\cal H} I.$$
The usual (smeared) {\it annihilation operator}, $\hat a(\a)$, is $(\hat
a^\dagger(C\a))^*$ where $C$ is the natural complex conjugation,
$\a\mapsto \a^*$ on $\cal H$.  Both of these operators annihilate the
{\it Fock vacuum vector} $\Omega^{\cal F}$. In this representation, the
one parameter group of time-translation automorphisms
\begin{equation} \label{auto}
\alpha(t): \sigma((\varphi_0, \pi_0); (f,p))\mapsto
\sigma((\varphi_t, \pi_t); (f,p))
\end{equation}
is implemented by $\exp(-iHt)$ where $H$ is the second quantization of
$\mu$ (i.e. the operator otherwise known as $\int\mu(k)\hat
a^\dagger(k)\hat a(k)d^3k$) on ${\cal F}({\cal H})$.

The most straightforward (albeit physically artificial) situation
involving ``particle creation'' in a curved spacetime concerns a
globally hyperbolic spacetime which, outside of a compact region, is
isometric to Minkowski space with a compact region removed -- i.e. to a
globally hyperbolic spacetime which is flat except inside a localized
``bump'' of curvature.  See Figure 1.  (One could also allow the
function $V$ in (\ref{kg}) to be non-zero inside the bump.)   On the
field algebra (defined as in the previous section) of such a spacetime,
there will be an ``in'' vacuum state (which may be identified with the
Minkowski vacuum to the past of the bump) and an ``out'' vacuum state
(which may be identified with the Minkowski vacuum to the future of the
bump) and one expects e.g. the ``in vacuum'' to arise as a many particle
state in the GNS representation of the ``out vacuum'' corresponding to
the creation of particles out of the vacuum by the bump of curvature.

\epsfclipon
\epsffile{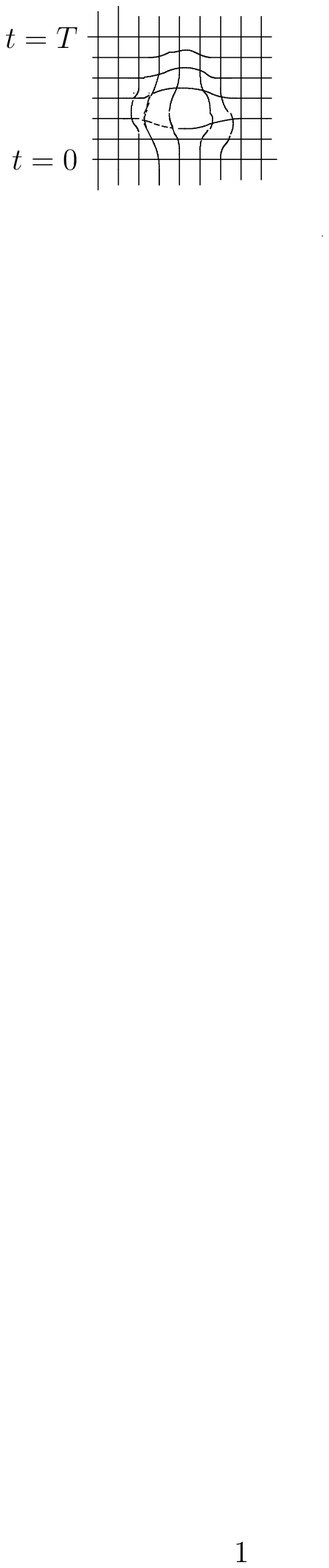}

\noindent
{\bf Figure 1} \quad A spacetime which is flat outside of a compact bump of
curvature.

\bigskip

In the formalism of this section, if we choose our global time
coordinate on such a spacetime so that, say, the $t=0$ surface is to the
past of the bump and the $t=T$ surface to its future, then the single
automorphism $\alpha(T)$ (defined as in (\ref{auto})) encodes the
overall effect of the bump of curvature on the quantum field and one can
ask whether it is implemented by a unitary operator in the GNS
representation of the Minkowski vacuum state (\ref{mink}).

This question may be answered by referring to the real-linear map ${\cal
T}: {\cal H}\rightarrow {\cal H}$ which sends
$\a_T=2^{-1/2}(\mu^{1/2}f_T+i\mu^{-1/2}p_T)$ to
$\a_0=2^{-1/2}(\mu^{1/2}f_0+i\mu^{-1/2}p_0)$.  By the conservation in
time of $\sigma$ and the symplecticity, noted in passing above, of the
map $K:(f,p)\mapsto \a$, this satisfies the defining relation
$$\rm{Im}\langle {\cal T} a^1|{\cal T} a^2\rangle=\rm{Im}\langle
a^1|a^2\rangle$$
of a {\it classical Bogoliubov transformation}.  Splitting ${\cal T}$
into its complex-linear and complex-antilinear parts by writing $${\cal
T} = \alpha + \beta C$$ where $\alpha$ and $\beta$ are complex linear
operators,  this relation may alternatively be expressed in terms of the
pair of relations $$\alpha^*\alpha-\bar\beta^*\bar\beta=I, \quad
\bar\alpha^*\bar\beta =  \beta^*\alpha$$ where $\bar\alpha=C\alpha C$,
$\bar\beta=C\beta C$.

We remark that there is an easy-to-visualize equivalent way of defining
$\alpha$ and $\beta$ in terms of the analysis, to the past of the bump,
into {\it positive and negative-frequency parts} of complex solutions to
(\ref{kg}) which are purely {\it positive-frequency} to the future of
the bump.  In fact, if, for any element $\a\in {\cal H}$, we identify
the positive frequency solution to the Minkowski-space Klein Gordon
equation
$$\phi^{\hbox{pos}}_{\hbox{out}}(t, {\bf x})=
((2\mu)^{-1/2}\exp(-i\mu t)\a)({\bf x})$$
with a complex solution to (\ref{kg}) to the future of the bump, then
(it may easily be seen) to the past of the bump, this same solution will
be identifiable with the (partly positive-frequency, partly
negative-frequency) Minkowski-space Klein-Gordon solution
$$\phi_{\hbox{in}}(t, {\bf x})=\left ((2\mu)^{-1/2}\exp(-i\mu t\right )
\alpha\a)({\bf x}) +
\left ((2\mu)^{-1/2}\exp(i\mu t)\bar\beta\a\right )({\bf x})$$
and this could be taken to be the defining equation for the operators $\alpha$
and
$\beta$.

It is then known (by a 1962 theorem of Shale) that the automorphism
(\ref{auto}) will be unitarily implemented if and only if $\beta$ is a
Hilbert Schmidt operator on $\cal H$.  Wald (1979, in case $m\ge 0$) and
Dimock (1979, in case $m\ne 0$) have verified that this condition is
satisfied in the case of our bump-of-curvature situation.  In that case,
if we denote the unitary implementor by $U$, we have the results

\begin{description}

\item{(i)} The expectation value $\langle U\Omega|N(\a)U\Omega\rangle_{{\cal
F}(\cal H)}$ of the number operator, $N(\a)=\hat a^\dagger(\a)\hat a(\a)$,
where $\a$ is a normalized element of ${\cal H}$, is equal to
$\langle\beta\a|\beta\a\rangle_{\cal H}$.

\item{(ii)} First note that there exists an orthonormal basis of
vectors,
$e_i$, ($i=1\dots\infty$), in $\cal H$ such that the (Hilbert
Schmidt) operator $\bar\beta^*\bar\alpha^{*-1}$ has the canonical form
$\sum_i \lambda_i \langle Ce_i|\cdot\rangle |e_i\rangle$.  We then have (up to
an
undertermined phase)
$$U\Omega=N\exp\left (-{1\over
2}\sum_i\lambda_i\hat a^\dagger(e_i)\hat a^\dagger(e_i)\right) \Omega.$$
where the normalization constant $N$ is chosen so that $\|U\Omega\|=1$
This formula makes manifest that the particles are created in pairs.

\end{description}

\noindent
We remark that, identifying elements, $\a$, of $\cal H$ with
positive-frequency solutions (below, we shall call them ``modes'') as
explained above, Result (i) may alternatively be expressed by saying
that the expectation value, $\omega_{\hbox{in}}(N(\a))$,  in the {\it
in-vacuum state} of the  occupation number, $N(\a)$, of a {\it
normalized mode}, $\a$, to the future of the bump is given by
$\langle\beta\a|\beta\a\rangle_{\cal H}$.

This formalism and the results, (i) and (ii) above, will generalize  (at
least heuristically, and sometimes rigorously -- see especially the
rigorous scattering theoretic work in the 1980's by Dimock and Kay and
more recently by A. Bachelot and others) to more realistic spacetimes
which are only asymptotically flat or asymptotically stationary.  In
favourable cases, one will still have notions of classical solutions
which are positive-frequency asymptotically towards the future/past,
and, in consequence, one will have well-defined asymptotic notions of
``vacuum'' and ``particles''.  Also, in, e.g. cosmological, models where
the background spacetime is slowly-varying in time, one can define
approximate {\it adiabatic} notions of classical positive frequency
solutions, and hence also of quantum ``vacuum'' and ``particles'' at
each finite value of the cosmological time.  But, at times where the
gravitational field is rapidly varying, one does not expect there to be
any sensible notion of ``particles''.  And, in a rapidly
time-varying background gravitational field which never settles down
one does not expect there to be any sensible particle interpretation of
the theory at all. To understand these statements, it suffices to consider the
$1+0$-dimensional Klein-Gordon equation with an external potential $V$:
$$\left (-{d^2\over dt^2}-m^2-V(t)\right )\phi=0$$ 
which is of course a system of one degree of freedom, mathematically
equivalent to the harmonic oscillator with a time-varying angular
frequency $\varpi(t)=(m^2+V(t))^{1/2}$. One could of course express its
quantum theory in terms of a time-evolving Schr\"odinger wave function
$\Psi(\varphi, t)$ and attempt to give this a particle interpretation at
each time, $s$, by expanding $\Psi(\varphi, s)$ in terms of the harmonic
oscillator wave functions for a harmonic oscillator with some particular
choice of angular frequency.  But the problem is, as is easy to convince
oneself, that there is no such good choice.   For example, one might
think that a good choice would be to take, at time $s$, the set of
harmonic oscillator wave functions with angular frequency $\varpi(s)$.
(This is sometimes known as the method of ``instantaneous
diagonalization of the Hamiltonian'').   But suppose we were to apply
this prescription to the case of a smooth $V(\cdot)$ which is constant
in time until time $0$ and assume the initial state is the usual vacuum
state. Then at some positive time $s$, the number of particles predicted
to be present is the same as the number of particles predicted to be
present on the same prescription at all times after $s$ for a $\hat
V(\cdot)$ which is equal to  $V(\cdot)$ up to time $s$ and then takes
the constant value $V(s)$ for all later times (see Figure 2).  But $\hat V(\cdot)$
will generically have a sharp corner in its graph (i.e. a discontinuity in
its time-derivative) at time $s$ and one would expect a large part of
the particle production in the latter situation to be accounted for by
the presence of this sharp corner -- and therefore a large part of the
predicted particle-production in the case of $V(\cdot)$ to be spurious.

\epsfclipon
\epsffile{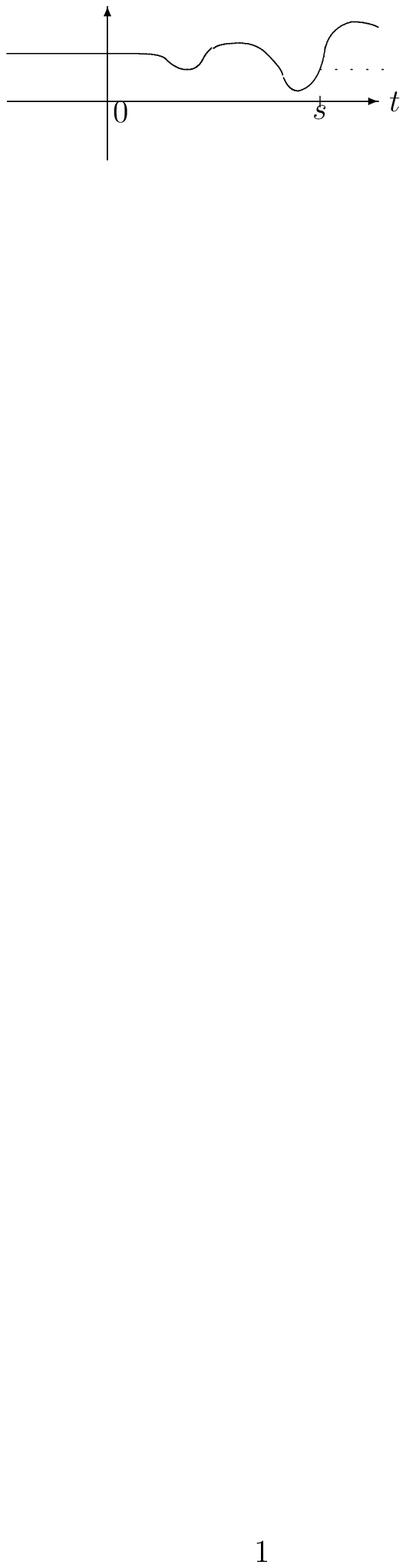}

\noindent
{\bf Figure 2} \quad Plots of $\varpi$ against $t$ for the two potentials
$V$ (continuous line) and $\hat V$ (continuous line up to $s$ and then dashed
line) which play a role in our critique of ``instantaneous diagonalization''.

\bigskip

Back in $1+3$ dimensions, even where a good notion of particles is
possible, it depends on the choice of time-evolution, as is dramatically
illustrated by the Unruh effect (see Section 5).

\section{Theory of the Stress-Energy Tensor}

To orient ideas, consider first the  free (minimally coupled) scalar
field,  $(\Box - m^2)\phi=0$, in Minkowski space.  If one quantizes this
system in the usual Minkowski-vacuum representation, then the
expectation value of the {\it renormalized stress-energy tensor}  (which
in this case is the same thing as the {\it normal ordered stress-energy
tensor}) in a vector state $\Psi$ in the Fock space will be given by the
formal {\it point-splitting} expression
$$\langle \Psi|T_{ab}(x)\Psi\rangle=\lim_{(x_1,x_2)\rightarrow (x,x)}
\left (\partial_a^1\partial_b^2-{1\over 2}\eta_{ab}(\eta^{cd}\partial_c^1
\partial_d^2+m^2)\right)$$
\begin{equation} \label{renvec}
\left(\langle\Psi|\rho_0(\phi(x_1)\phi(x_2))\Psi\rangle
-\langle\Omega^{\cal F}|\rho_0(\phi(x_1)\phi(x_2))\Omega^{\cal
F}\rangle\right)
\end{equation}
where $\eta_{ab}$ is the usual Minkowski metric. A sufficient condition
for the limit here to be finite and well-defined would, e.g., be for
$\Psi$ to consist of a (normalised) finite superposition of $n$-particle
vectors of form $\hat a^\dagger(\a_1),\dots ,\hat
a^\dagger(\a_n)\Omega^{\cal F}$ where the smearing functions $\a_1,\dots
,\a_n$ are all $C^\infty$ elements of $\cal H$ (i.e. of $L^2_{\bf C}({\bf
R}^3)$.  The reason this works is that the two-point function in such
states shares the same short-distance singularity as the
Minkowski-vacuum two-point function.  For exactly the same reason, one
obtains a well-defined finite limit if one defines the expectation value
of the stress-energy tensor in any physically admissible quasi-free
state by the expression
\begin{equation} \label{ren}
\omega(T_{ab})(x))=\lim_{(x_1,x_2)\rightarrow (x,x)}
\left (\partial_a^1\partial_b^2-{1\over 2}\eta_{ab}(\eta^{cd}\partial_c^1
\partial_d^2+m^2)\right)\left(\omega(\phi(x_1)\phi(x_2))
-\omega_0(\phi(x_1)\phi(x_2))\right ).
\end{equation}
This latter point-splitting formula generalizes to a definition for the
{\it expectation value of the renormalized stress-energy tensor} for an
arbitrary physically admissible quasi-free state (or indeed for an
arbitrary state whose two point function has {\it Hadamard form} -- i.e.
whose anticommutator function satisfies Condition (d) of Section 2) on
the minimal field algebra and to other linear field theories (including
the stress tensor for a conformally coupled linear scalar field) on a
general globally hyperbolic spacetime (and the result obtained agrees
with that obtained by other methods, including {\it dimensional
regularization} and {\it zeta-function regularization}).  However, the
generalization to a curved spacetime involves a number of important new
features which we now briefly list. (See (Wald, 1978) for details.)

First, the subtraction term which replaces
$\omega_0(\phi(x_1)\phi(x_2))$ is, in general,  not the expectation
value of $\phi(x_1)\phi(x_2)$ in any particular state, but rather a
particular {\it locally constructed Hadamard two-point function} whose
physical interpretation is more subtle; the renormalization is thus in
general not to be regarded as a normal ordering.  Second, the immediate
result of the resulting limiting process will not be covariantly
conserved and, in order to obtain a covariantly conserved quantity, one
needs to add a particular {\it local geometrical correction term}.  The
upshot of this is that the resulting expected stress-energy tensor is
covariantly conserved but possesses a (state-independent) {\it anomalous
trace}.  In particular, for a massless conformally coupled linear scalar
field, one has (for all physically admissible quasi-free states,
$\omega$) the {\it trace anomaly formula}
$$\omega(T^a_a(x))=(2880\pi^2)^{-1}(C_{abcd}C^{abcd}+R_{ab}R^{ab}-{1\over
 3}R^2)$$

 -- plus an arbitrary multiple of $\Box R$.  In fact, in general, the
thus-defined renormalized stress-energy tensor operator (see below) is
only defined up to a {\it finite renormalization ambiguity} which
consists of the addition of arbitrary multiples of the functional
derivatives with respect to $g_{ab}$ of the four quantities
$$I_n=\int_{\cal M} F_n(x)|\det(g)|^{1/2}d^4x,$$
$n=1\dots 4$, where $F_1=1$, $F_2=R$, $F_3=R^2$, $F_4=R_{ab}R^{ab}$.   In the
Minkowski-space case, only the first of these ambiguities arises and it
is implicitly resolved in the formulae (\ref{renvec}), (\ref{ren})
inasmuch as these effectively incorporate the {\it renormalization
condition} that $\omega_0(T_{ab})=0$.  (For the same reason, the
locally-flat example we give below has no ambiguity.)

One expects, in both flat and curved cases, that, for test functions,
$F\in C_0^\infty({\cal M})$, there will exist operators $T_{ab}(F)$
which are {\it affiliated to} the net of local  $W^*$-algebras referred
to in Section 2 and that it is meaningful to write 
$$\int_{\cal
M}\omega(T_{ab}(x))F(x)|\det(g)|^{1/2}d^4x=\omega(T_{ab}(F))$$  
provided that, by $\omega$ on the right-hand side, we understand the
extension of $\omega$ from the Weyl algebra to this net.  ($T_{ab}(F)$
is however not expected to belong to the minimal algebra or be
affiliated to the Weyl algebra.)

An interesting simple example of a renormalized stress-energy tensor
calculation is the so-called {\it Casimir effect} calculation for a
linear scalar field on a (for further simplicity, $1+1$-dimensional)
{\it timelike cylinder spacetime} of radius $R$ (see Figure 3). This
spacetime is globally hyperbolic and stationary and, while locally flat,
globally distinct from Minkowski space.  As a result, while  -- provided
the regions ${\cal O}$ are sufficiently small (such as the diamond
region in Figure 3) -- elements ${\cal A}({\cal O})$ of the minimal net
of local algebras on this spacetime will be identifiable, in an obvious
way, with elements of the minimal net of local algebras on Minkowski
space, the stationary ground state $\omega_{\hbox{cylinder}}$ will, when
restricted to such thus-identified regions, be distinct from the
Minkowski vacuum state $\omega_0$.  The resulting renormalized
stress-energy tensor (as first pointed out in (Kay, 1979), defineable,
once the above identification has been made, exactly as in (\ref{ren}))
turns out to be non-zero and, interestingly, to have a (in the natural
coordinates, constant) {\it negative energy density} $T_{00}$.  In fact:
$$\omega_{\hbox{cylinder}}(T_{ab})={1\over 24\pi R^2} \eta_{ab}.$$

\epsfclipon
\epsffile{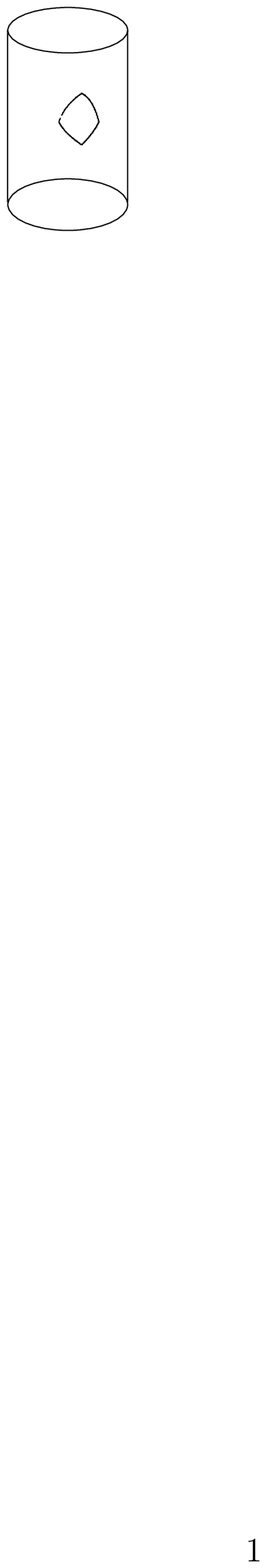}

\noindent
{\bf Figure 3}\quad The time-like cylinder spacetime of radius $R$ with a
diamond region isometric to a piece of Minkowski space.  See (Kay, 1979).

\section{Hawking and Unruh effects}

The original (1974) calculation by Hawking concerned a model spacetime
for a star which collapses to a black hole. For simplicity, we shall
only discuss the spherically-symmetric case.  (See Figure 4.) 

\epsfclipon
\epsffile{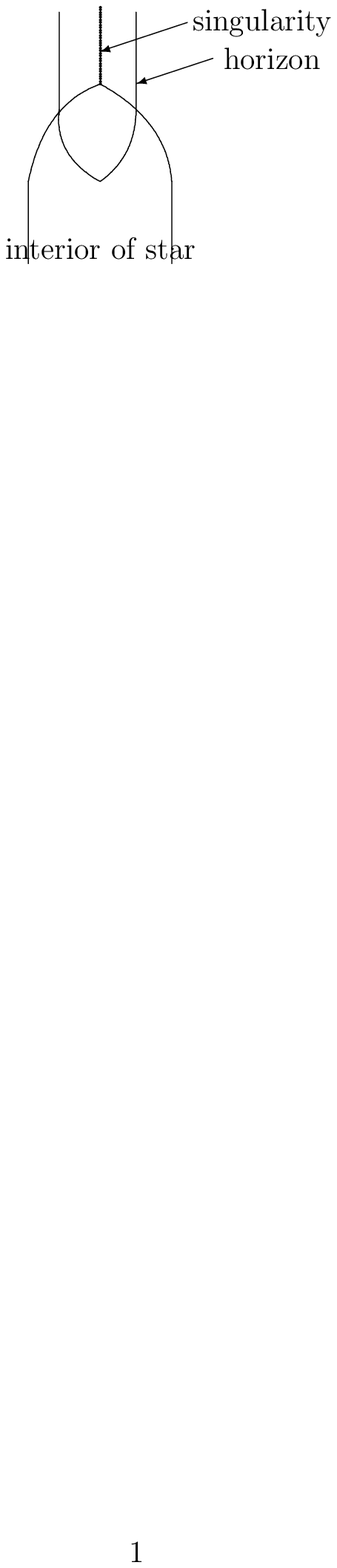}

\noindent
{\bf Figure 4}\quad The spacetime of a star collapsing to a spherical black
hole.

\bigskip

Adopting a similar ``mode'' viewpoint to that mentioned after Results
(i) and (ii) in Section 3, the result of the calculation may be stated
as follows: For a real linear scalar field satisfying (\ref{kg}) with
$m=0$ (and $V=0$) on this spacetime, the expectation value
$\omega_{\hbox{in}}(N(\a_{\varpi,\ell}))$ of the occupation number of a
one-particle outgoing mode $\a_{\varpi, \ell})$ localized (as far as a
normalized mode can be) around $\varpi$ in angular-frequency-space and
about retarded time $v$ and with angular momentum ``quantum number''
$\ell$, in the in-vacuum state (i.e. on the minimal algebra for a real
scalar field on this model spacetime)  $\omega_{\hbox{in}}$ is, at late
retarded times, given by the formula
$$\omega_{\hbox{in}}(N(\a_{\varpi,\ell}))={\Gamma(\varpi,\ell)\over
\exp(8\pi M\varpi)-1}$$
where $M$ is the mass of the black hole and the {\it absorption factor}
(alternatively known as {\it grey body factor})  $\Gamma(\varpi, \ell)$
is equal to the norm-squared of that part of the one-particle mode,
$\a_{\varpi,\ell}$, which, viewed as a complex positive frequency
classical solution propagating backwards in time from late retarded
times, would be absorbed by the black hole.  This calculation can be
understood as an application of Result (i) of Section 3 (even though the
spacetime is more complicated than one with a localized ``bump of
curvature'' and even though the relevant overall time-evolution will not
be unitarily implemented, the result still applies when suitably
interpreted) and the heart of the calculation is an asymptotic estimate
of the relevant ``$\beta$'' Bogoliubov coefficient which turns out to be
dependent on the geometrical optics of rays which pass through the star
just before the formation of the horizon.   This result suggests that
the in-vacuum state is indistinguishable at late retarded times from a
state of black-body radiation at the {\it Hawking temperature},
$T_{\hbox{Hawking}}=1/8\pi M$, in Minkowski space from a black-body with
the same absorption factor. This was confirmed by further work by many
authors. Much of that work, as well as the original result of Hawking
was partially heuristic but later work by Dimock and Kay (1987), by
Fredenhagen and Haag (1990) and by Bachelot (1999) and  others has put
different aspects of it on a rigorous mathematical footing.  The result
generalizes to non-zero mass and higher spin fields and to interacting
fields as well as to other types of black hole and the formula for the 
Hawking temperature generalizes to
$$T_{\hbox{Hawking}}=\kappa/2\pi$$
where $\kappa$ is the {\it surface gravity} of the black hole.

This result suggests that there is something fundamentally ``thermal''
about quantum fields on black-hole backgrounds and this is confirmed by
a number of mathematical results.  In particular, the theorems in the
two papers (Kay and Wald, 1991) and (Kay, 1993), combined together, tell
us that there is a unique state on the Weyl algebra for the {\it
maximally extended Schwarzschild spacetime} (a.k.a. {\it
Kruskal-Szekeres spacetime}) (see Figure 5) which is invariant under the
{\it Schwarzschild isometry group} and whose two-point function has
Hadamard form. Moreover, they tell us that this state, when restricted
to a single wedge (i.e. the exterior Schwarzschild spacetime) is
necessarily a KMS state at the Hawking temperature.  

\epsfclipon
\epsffile{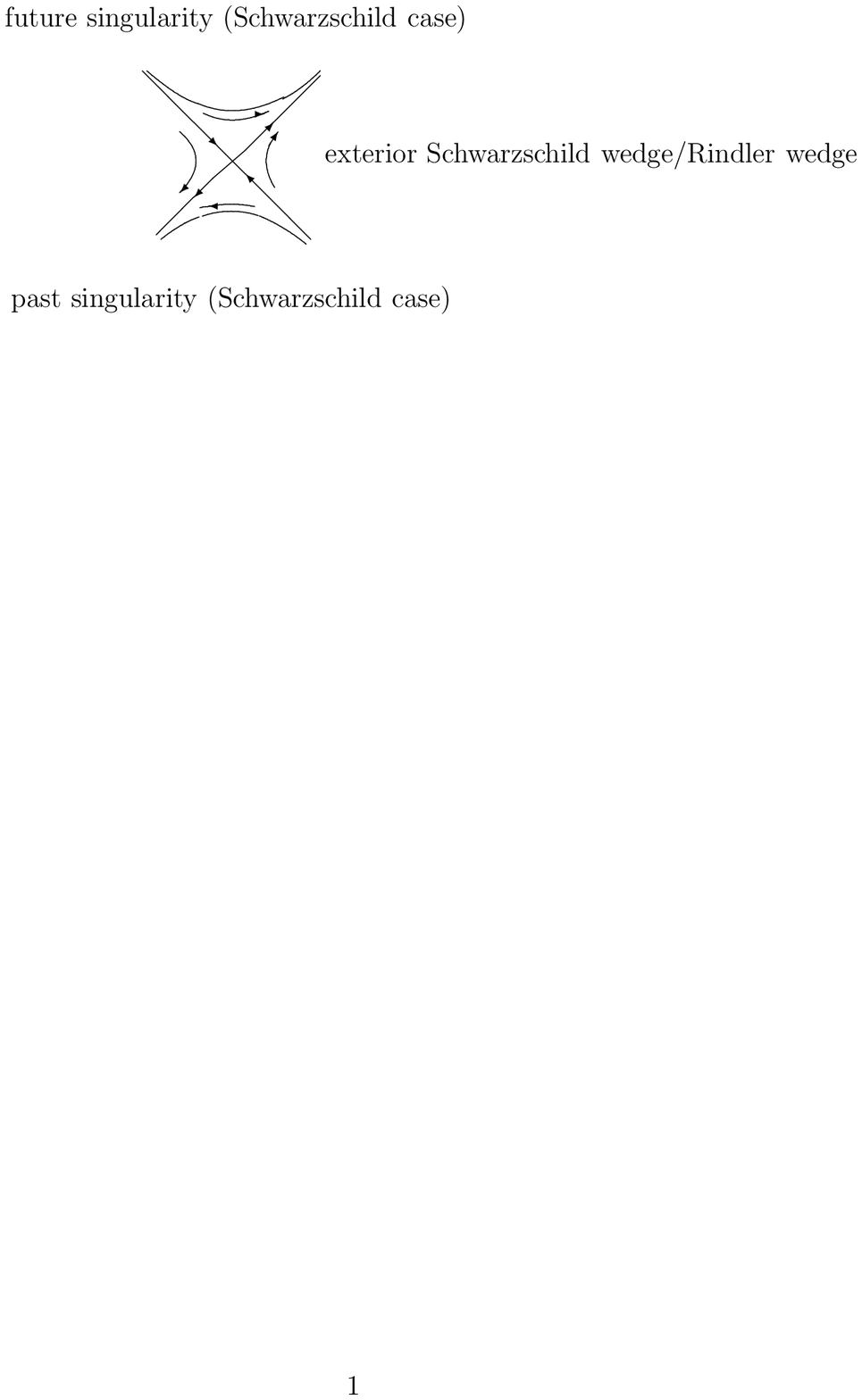}

\noindent
{\bf Figure 5}\quad The geometry of maximally extended Schwarzschild
(/or Minkowski) spacetime.  In the Schwarzschild case, every point
represents a two-sphere (/in the Minkowski case, a two-plane).  The
curves with arrows on them indicate the Schwarzschild time-evolution
(/one-parameter family of Lorentz boosts).  These curves include the (straight
lines at right angles) event horizons (/Killing horizons).

\bigskip

This unique state is known as the {\it Hartle-Hawking-Israel state}.
These results in fact apply more generally to a wide class of globally
hyperbolic spacetimes with {\it bifurcate Killing horizons} including de
Sitter space -- where the unique state is sometimes called the {\it
Euclidean}  and sometimes the {\it Bunch-Davies} vacuum state -- as well
as to Minkowski space, in which case the unique state is the usual
Minkowski vacuum state,  the analogue of the exterior Schwarzschild
wedge is a so-called {\it Rindler wedge}, and the relevant isometry
group is a one-parameter family of wedge-preserving Lorentz boosts. In
the latter situation, the fact that the Minkowski vacuum state is a KMS
state (at ``temperature'' $1/2\pi$) when restricted to a Rindler wedge
and regarded with respect to the time-evolution consisting of the
wedge-preserving one-parameter family of Lorentz boosts is known as the
{\it Unruh effect} (1975).  This latter property of the Minkowski vacuum
in fact generalizes to general {\it Wightman quantum field theories} and
is in fact an immediate consequence of a combination of the {\it Reeh
Schlieder Theorem} (applied to a Rindler wedge) and the {\it Bisognano
Wichmann Theorem} (1975).  The latter theorem says that the defining
relation (\ref{kms}) of a KMS state holds if, in (\ref{kms}), we
identify the operator $J$ with the complex conjugation which implements
wedge reflection and $H$ with the self-adjoint generator of the unitary
implementor of Lorentz boosts. We remark that the Unruh effect
illustrates how the concept of ``vacuum'' (when meaningful at all) is
dependent on the choice of time-evolution under consideration.  Thus the
usual Minkowski vacuum is a ground state with respect to the usual
Minkowski time-evolution but not (when restricted to a Rindler wedge)
with respect to a one-parameter family of Lorentz boosts; with respect
to these, it is, instead, a KMS state.

\section{Non-Globally Hyperbolic Spacetimes and the ``Time Machine'' Question}

In (Hawking, 1992) it is argued that a spacetime in which a time-machine gets
manufactured should be modelled (see Figure 6) by a spacetime with an
{\it initial globally hyperbolic region} with a region containing {\it
closed timelike curves} to its future and such that the future boundary
of the globally hyperbolic region is a {\it compactly generated Cauchy
horizon}. On such a spacetime, (Kay, Radzikowski and Wald, 1997) proves
that it is impossible for any distributional bisolution which
satisfies (even a certain weakened version of) the Hadamard condition on
the initial globally hyperbolic region to continue to satisfy that
condition on the full spacetime -- the (weakened) Hadamard condition
being necessarily violated at at least one point of the Cauchy horizon.
This result implies that, however one extends a state, satisfying our
conditions (a), (b), (c) and (d), on the minimal algebra for
({\ref{kg}}) on the initial globally hyperbolic region, the expectation
value of its stress-energy tensor must necessarily become singular {\it
on} the Cauchy horizon.  This result, together with many heuristic
results and specific examples considered by many other authors appears
to support the validity of the (Hawking, 1992) {\it chronology protection
conjecture} to the effect that it is impossible in principle to
manufacture a time machine. However, there are potential loopholes in
the physical interpretation of this result as pointed out by Visser
(1997), as well as other claims by various authors that one can
nevertheless violate the chronology protection conjecture.  For a recent
discussion on this question, we refer to (Visser, 2003).

\def\epsfsize#1#2{0.8#1}
\epsfclipon
\epsffile{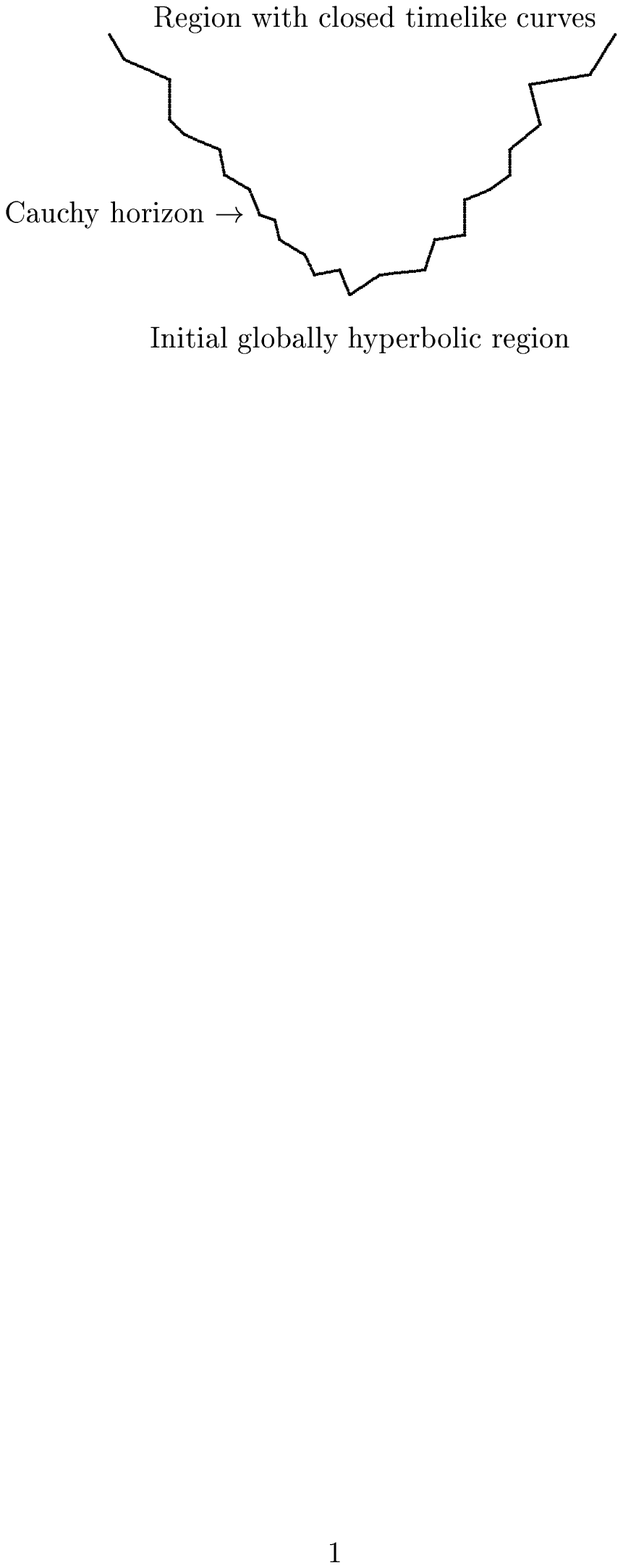}

\noindent
{\bf Figure 6}\quad The schematic geometry of a spacetime in which a
time-machine gets manufactured.

\section{Other Related Topics and Some Warnings}

There is a vast computational literature, calculating the expectation
values of stress-energy tensors in states of interest for scalar and
higher spin linear fields (and also some work for interacting fields) on
interesting cosmological and black-hole backgrounds.  Quantum field
theory on de Sitter and anti-de Sitter space is a big subject area in
its own right with recent renewed interest because of its relevance to
{\it string theory} and {\it holography}.   Also important on black hole
backgrounds is the calculation of grey-body factors, again with renewed
interest because of relevance to string theory and to {\it
brane-world} scenarios.

There are many further mathematically rigorous results on algebraic and
axiomatic quantum field theory in a curved spacetime setting, including
versions of {\it PCT}, {\it Spin-Statistics} and {\it Reeh-Schlieder}
theorems and also rigorous {\it energy inequalities} bounding the extent
to which expected energy densities can be negative etc.

There is much mathematical work controlling scattering theory on black
holes, partly with a view to further elucidating the Hawking effect.

Perturbative renormalization theory of interacting quantum fields in
curved spacetime is also now a highly developed subject.

Beyond quantum field theory in a fixed curved spacetime is {\it
semiclassical gravity} which takes into account the {\it back reaction}
of the expectation value of the stress-energy tensor on the classical
gravitational background.  There are also interesting condensed-matter
analogues of the Hawking effect such as {\it dumb holes}.

Readers exploring the wider literature, or doing further research on the
subject should be  aware that the word ``vacuum'' is sometimes used to
mean ``ground state'' and sometimes just to mean ``quasi-free state''. 
Furthermore they should be cautious of attempts to define
particles on Cauchy surfaces in {\it instantaneous diagonalization}
schemes (cf. the remarks at the end of Section 3).   When studying (or
performing) calculations of the ``expectation value of the stress-energy
tensor'' it is always important to ask oneself with respect to which
state the expectation value is being taken.  It is also important to
remember to check that candidate two-point (anticommutator) functions
satisfy the positivity condition (c) of Section 2.  Typically two-point
distributions obtained via {\it mode sums} automatically satisfy
Condition (c) (and Condition (d)), but those obtained via {\it image}
methods don't always satisfy it.  (When they don't, the presence of
non-local spacelike singularities is often a tell-tale sign as can be
inferred from Kay's Conjecture/Radzikowski's Theorem discussed in Section
2.) There are a number of apparent implicit assertions in the literature
that some such two-point functions arise from ``states'' when of course
they can't.  Some of these concern proposed analogues to the
Hartle-Hawking-Israel state for the (appropriate maximal globally hyperbolic
portion of the maximally extended) Kerr spacetime.  That they can't
belong to states is clear from  a theorem in (Kay and Wald, 1991) which
states that there is no stationary Hadamard state on this spacetime at
all.   Others of them concern claimed ``states'' on spacetimes such as
those discussed in Section 6 which, if they really were states would
seem to be in conflict with the chronology protection conjecture. 
Finally, beware states  (such as the so-called {\it $\alpha$-vacua} of
de Sitter spacetime) whose two-point distributions violate the
``Hadamard'' Condition (d) of Section 2 and which therefore do not have
a well-defined finite expectation value for the renormalized
stress-energy tensor.

\section{See also}

{\bf Black Hole Thermodynamics}. {\bf Algebraic Approach to Quantum Field
Theory}.

\section{Further Reading}

\begin{description}

\item DeWitt BS (1975) Quantum field theory in curved space-time.
{\it Physics Reports} 19, No. 6, 295-357.

\item Birrell ND and Davies PCW (1982) {\it Quantum Fields in Curved Space.}
Cambridge: Cambridge University Press.

\item Haag R (1996) {\it Local Quantum Physics.}  Berlin: Springer.

\item Dimock J (1980) Algebras of local observables on a manifold. {\it
Communications in Mathematical Physics} 77: 219-228.

\item Brunetti R, Fredenhagen K and Verch R (2003)
The generally covariant locality principle -- A new paradigm for local quantum
physics.
{\it Communications in Mathematical Physics} 237:31-68.

\item Kay, BS (2000) Application of linear hyperbolic PDE to linear quantum
fields in curved spacetimes: especially black holes, time machines and a
new semi-local vacuum concept.  Journ\'ees {\it \'Equations aux
D\'eriv\'ees Partielles},  Nantes, 5-9 juin 2000 GDR 1151 (CNRS): IX1-IX19.
(Also available at http://www.math.sciences.univ-nantes.fr/edpa/2000/html
or as gr-qc/0103056.)

\item Wald RM (1978) Trace anomaly of a conformally invariant
quantum field in a curved spacetime.  {\it Physical Review} D17: 1477-1484.

\item Kay BS (1979) Casimir effect in quantum field theory. ({\it
Original title:} The Casimir effect without magic.) {\it
Physical Review} D20: 3052-3062.

\item Hawking SW (1975) Particle Creation by Black Holes.  {\it
Communications in Mathematical Physics} 43: 199-220.

\item Kay BS and Wald RM (1991) Theorems on the uniqueness and
thermal properties of stationary, nonsingular, quasifree states on
spacetimes with a bifurcate Killing horizon. {\it Physics Reports} 207,
No. 2, 49-136.

\item Kay BS (1993) Sufficient conditions for quasifree states and an
improved uniqueness theorem for quantum fields on space-times with
horizons. {\it Journal of Mathematical Physics} 34: 4519-4539.

\item Wald RM (1994) {\it Quantum Field Theory in Curved Spacetime and Black
Hole Thermodynamics.} Chicago: University of Chicago Press.

\item Hartle JB and Hawking SW (1976) Path-integral derivation of
black-hole radiance.  {\it Physical Review} D13: 2188-2203.

\item Israel W (1976) Thermo-field dynamics of black holes.
{\it Physics Letters A} 57: 107-110.

\item Unruh W (1976) Notes on black hole evaporation.  {\it Physical
Review} D14: 870-892.

\item Hawking SW (1992) The Chronology Protection Conjecture.
{\it Physical Review} D46: 603-611.

\item Kay BS, Radzikowski MJ and Wald RM (1997) Quantum field theory
on spacetimes with a compactly generated Cauchy horizon.  {\it
Communications in Mathematical Physics} 183: 533-556.

\item Visser M (2003) The quantum physics of chronology protection.  In:
Gibbons GW, Shellard EPS and Rankin SJ {\it The Future of Theoretical Physics
and Cosmology.} Cambridge: Cambridge University Press.

\end{description}

\end{document}